\documentclass[12pt]{article}
\usepackage{multirow}  
\usepackage{amsmath}
\usepackage{amssymb, latexsym}
 \usepackage{amsthm}
 \usepackage{caption}
\usepackage{setspace} 
\renewcommand{\baselinestretch}{1}
\def\theequation{\thesection.\arabic{equation}}
\def\no{\noindent}
\def\mS{\mathcal{S}}
\def\mU{\mathcal{U}}
\def\N{\scriptscriptstyle N}

\def\A{\scriptscriptstyle A}
\def\B{\scriptscriptstyle B}

\def\REG{\scriptscriptstyle REG}
\def\SM{\scriptscriptstyle SM}
\def\IPWone{\scriptscriptstyle IPW1}
\def\IPWtwo{\scriptscriptstyle IPW2}
\def\DR{\scriptscriptstyle DR}
\def\PLUG{\scriptscriptstyle PLUG}
\def\KH{\scriptscriptstyle KH}

\def\Cone{\scriptscriptstyle C1}
\def\Ctwo{\scriptscriptstyle C2}

\def\DR{\scriptscriptstyle DR}
\def\DRone{\scriptscriptstyle DR1}
\def\DRtwo{\scriptscriptstyle DR2}

\def\PRC{\scriptscriptstyle PRC}
\def\BRFSS{\scriptscriptstyle BRFSS}
\def\CPS{\scriptscriptstyle CPS}

\def\bfx{\mbox{\boldmath{$x$}}}

\def\bfu{\mbox{\boldmath{$u$}}}

\def\bftheta{\mbox{\boldmath{$\theta$}}}
\def\bfbeta{\mbox{\boldmath{$\beta$}}}

\def\bfeta{\mbox{\boldmath{$\eta$}}}
\def\bfzero{\mbox{\boldmath{$0$}}}
\def\bfphi{\mbox{\boldmath{$\phi$}}}
\def\bfPhi{\mbox{\boldmath{$\Phi$}}}

\def\mU{\mathcal{U}}

\def\mS{\mathcal{S}}
\def\mF{\mathcal{F}}

\begin{document}

\centerline{\large {\bf Doubly Robust Inference with Non-probability Survey Samples}}

\bigskip

\centerline{Yilin Chen, \ Pengfei Li \ and \ Changbao Wu\footnote{Yilin Chen is doctoral student, Pengfei Li is Associate Professor and Changbao Wu is Professor, Department of Statistics and Actuarial Science, University of Waterloo, Waterloo ON N2L 3G1,Canada (E-mails: {\em yilin.chen@uwaterloo.ca}, \ {\em pengfei.li@uwaterloo.ca} \ and \ {\em cbwu@uwaterloo.ca}). This research was supported by grants from the Natural Sciences and Engineering Research Council of Canada. The authors thank Drs. Mike Brick of Westat, Courtney Kennedy and Andrew Mercer of Pew Research Center for assistance with the PEW dataset.}}

\bigskip

\bigskip

\hrule

{\small
\begin{quotation}
\no
We establish a general framework for statistical inferences with non-probability survey samples when relevant auxiliary information is available from a probability survey sample. We develop a rigorous procedure for estimating the propensity scores for units in the non-probability sample, and construct doubly robust estimators for the finite population mean. Variance estimation is discussed under the proposed framework. Results from simulation studies show the robustness and the efficiency of our proposed estimators as compared to existing methods. The proposed method is used to analyze a non-probability survey sample collected by the Pew Research Center with auxiliary information from the Behavioral Risk Factor Surveillance System and the Current Population Survey. Our results illustrate a general approach to inference with non-probability samples and highlight the importance and usefulness of auxiliary information from probability survey samples.

\vspace{0.3cm}

\no
KEY WORDS \ Design-based inference, inclusion probability, missing at random, probability sampling, propensity score, regression modeling, variance estimation. 
\end{quotation}
}

\hrule

\bigskip

\bigskip

\setcounter{section}{1}
\setcounter{equation}{0}
\no {\bf 1. INTRODUCTION}

\medskip

\no
Probability sampling methods have become a universally accepted approach in survey sampling since the seminal paper of Neyman (1938). Design-based inferences for finite populations using probability survey samples are widely adopted by official statistics and researchers in areas such as social studies and health sciences where surveys are one of the primary tools for data collection (Hansen 1987; Rao, 2005). There exists an extensive literature with continued research activities on probability sampling and design-based inferences for complex surveys. 

The use of non-probability survey samples has a very long history. Quota sampling, for instance, serves as a cost-effective alternative method to select a survey sample when one is limited by resources and/or the availability of reliable sampling frames. However, non-probability sampling methods have never gained true  momentum in survey practice of the 20th century due to the lack of theoretical foundation for statistical inferences. 

The success of probability sampling has led to more frequent surveys and more ambitious research projects that involve long and sophisticated questionnaires and measurements. Response burden and privacy concerns, along with many other factors, have led to dramatic decease in response rates for almost all surveys. The challenge of low participation rates and the ever-increasing costs for conducting surveys using probability sampling methods, coupled with technology advances,  has resulted in a shift of paradigm in recent years for government  agencies, research institutions and industrial organizations to seek other cheaper and quicker alternatives for data collection (Citro, 2014). In particular, a great deal of attention has been given to non-probability survey samples.

The rise of the web based surveys has reshaped our views on non-probability sampling in terms of cost-and-time efficiency. The most popular type of web surveys is based on the so-called {\em opt-in panels}. These panels consist of volunteers who agreed to participate and are recruited through various convenient but non-probability methods. Online research through opt-in panel surveys has become popular in recent years due to its efficient recruitment process, quick responses, and low maintenance expenses. Tourangeau et al. (2013) contains many examples for web based surveys. 

As much as the excitement brought by these changes, there are serious issues and major challenges for the use of web surveys and other non-probability survey samples. The ``{\em Summary Report of the AAPOR Task Force on Non-probability Sampling}'' by Baker et al. (2013), which was commissioned by the American Association of Public Opinion Research (AAPOR) Executive Council, contains a well documented list of such issues and challenges. The task force's conclusions include: (i) unlike probability sampling, there is no single framework that adequately encompasses all of non-probability sampling; (ii) making inferences for any probability or non-probability survey requires some reliance on modeling assumptions; and (iii) if non-probability samples are to gain wider acceptance among survey researchers there must be a more coherent framework and accompanying set of measures for evaluating their quality. 

In this paper, we propose a general framework for statistical inferences with non-probability survey samples when relevant auxiliary information is available from a reference probability survey sample. Our major contributions are the development of a rigorous procedure for estimating the propensity scores for units in the non-probability sample and the subsequent estimation procedures for parameters of the finite population. It is shown that the doubly robust estimators developed in the context of missing data and causal inference can be conveniently used for inferences with non-probability samples. 
 
The rest of the paper is organized as follows. Section 2 introduces notation and describes the issue of estimating the propensity scores and one of the popular inferential procedures on sample matching with non-probability samples. Section 3 presents the proposed procedure for estimating the propensity scores and the doubly robust estimator for the finite population mean of the response variable. Variance estimation for the proposed estimators is discussed in Section 4. Results of simulation studies on the finite sample performance of the proposed estimators with comparisons to existing estimators are presented in Section 5. An application of the proposed method to analyzing a non-probability survey sample collected by the Pew Research Center with auxiliary information from the Behavioral Risk Factor Surveillance System and the Current Population Survey is presented in Section 6. Some concluding remarks are given in Section 7. Proofs and technical details are given in the Appendix.

\bigskip

\setcounter{section}{2}
\setcounter{equation}{0}
\no {\bf 2. PROPENSITY SCORES AND SAMPLE MATCHING}

\medskip

\no
The major hurdle in analyzing data from non-probability survey samples is the unknown selection mechanism for a unit to be included in the sample. It belongs to the so-called ``biased sample'' problem and the sample itself does not provide a valid picture about the entire finite population. Our proposed approach is based on the assumption that relevant auxiliary information is available from a reference probability survey sample. This setting has previously been used by several authors. See, for instance, Rivers (2007), Vavreck and Rivers (2008), Lee and Valliant (2009) and Brick (2015). 

\medskip

\no {\bf 2.1  Basic Setting and Propensity Scores}

\medskip

\no
Let $\mU=\{1,2,\cdots,N\}$ represent the set of $N$ units for the finite population, with $N$ being the population size. Associated with unit $i$ are values of the vector of auxiliary variables, $\bfx_i$, and the value $y_i$ for the response variable $y$, $i = 1, 2, \cdots, N$. Under the design-based framework, the set of finite population values $\mF_{\N} = \{(\bfx_i,y_i),i\in \mU\}$ is viewed as fixed. Let $\mu_y = N^{-1}\sum_{i=1}^N y_i$ be the finite population mean for the response variable. 

Consider a non-probability sample $\mS_{\A}$ consisting of $n_{\A}$ units from the finite population. Let $\{(\bfx_i,y_i),i \in \mS_{\A}\}$ be the dataset from the non-probability sample. Let $R_i=I(i\in \mS_{\A})$ be the indicator variable for unit $i$ being included in the sample $\mS_{\A}$, i.e., $R_i=1$ if $i\in \mS_{\A}$ and $R_i=0$ if $i \notin \mS_{\A}$, $i=1,2,\cdots,N$. The propensity scores are given by 
\[
\pi_i^{\A} = E_q(R_i \mid \bfx_i, y_i) = P_q(R_i = 1 \mid \bfx_i, y_i)\,, \;\; i=1,2,\cdots,N\,,
\]
where the subscript $q$ refers to the model for the selection mechanism for the sample $\mS_{\A}$, i.e., the propensity score model. 

The selection mechanism is called {\em ignorable} if $\pi_i^{\A} = P(R_i = 1 \mid \bfx_i, y_i) = P(R_i = 1 \mid \bfx_i)$ for all $i$. This corresponds to {\em missing at random} (MAR) as defined by Rubin (1976) and Little and Rubin (2002). More formally, we consider the following assumptions for the selection mechanism. 

\begin{itemize}
\item[{\bf A1}] \ The selection indicator $R_i$ and the response variable $y_i$ are independent given the set of covariates $\bfx_i$. 
\item[{\bf A2}] \ All units have a non-zero propensity score, i.e., $\pi_i^{\A} >0$ for all $i$. 
\item[{\bf A3}] \ The indicator variables $R_i$ and $R_j$ are independent given $\bfx_i$ and $\bfx_j$ for $i\neq j$. 
\end{itemize}

As pointed out by Rivers (2007), the term ``ignorable'' is an unfortunate choice of terminology for the missing data and causal inference literature, since it certainly cannot be ignored by the analyst. Similarly, the term ``missing at random'' should not be confused with ``randomly missing''. Assumption {\bf A2} cannot be satisfied by scenarios where certain units will for sure not be included in the sample. This is a rather complicated issue which will not be pursued further in the paper. Assumptions {\bf A1} and {\bf A2} together are the strong ignorability condition as discussed by Rosenbaum and Rubin (1983). 

The propensity scores $\pi_i^{\A}$ cannot be estimated from the sample $\mS_{\A}$ itself, and information on the rest of the finite population is required. Suppose that a reference probability sample, denoted as $\mS_{\B}$, with auxiliary information on $\bfx$ is available as an existing survey or can be obtained without much difficulty. Let $\{(\bfx_i,d_i^{\B}), i \in \mS_{\B}\}$ be the data from the reference probability sample, where $d_i^{\B}=1/\pi_i^{\B}$ are the survey weights and $\pi_i^{\B} = P(i\in \mS_{\B})$ are the inclusion probabilities under the probability sampling design for the sample $\mS_{\B}$. Note that the response variable $y$ is not part of the dataset for the reference sample. 

Existing methods for estimating the propensity scores do not provide valid results. The approaches described in Rivers (2007), Lee (2006), Lee and Valliant (2009) and Brick (2015) attempt to estimate $\tilde{\pi}_i^{\A} = P(\tilde{R}_i =1 \mid \bfx_i)$, where $\tilde{R}_i =1$ if $i\in \mS_{\A}$ and $\tilde{R}_i =0$ if $i\in \mS_{\B}$. It is obvious that pooling the two samples $\mS_{\A}$ and $\mS_{\B}$ together in such a way does not provide the required information for the estimation of the true propensity scores $\pi_i^{\A} = P(R_i =1 \mid \bfx_i)$. The resulting estimators using the estimated propensity scores $\tilde{\pi}_i^{\A}$ are biased. This has been previously observed by Valliant and Dever (2011). Our proposed rigorous estimation procedure is presented in Section 3.1.

\medskip

\no {\bf 2.2  Sample Matching and the Prediction Approach}

\medskip

\no
Model-based prediction approach has been explored for inferences with non-probability samples. Suppose that the finite population $\{(\bfx_i,y_i),i\in \mU\}$ can be viewed as a random sample from the model 
\begin{equation}
\label{reg}
y_i = m(\bfx_i) + \varepsilon_i\,, \;\; i=1,2,\cdots,N\,,
\end{equation}
where $m(\bfx_i)=E_\xi(y_i \mid \bfx_i)$, which can take a parametric form such as $m(\bfx_i)=\bfx_i^{\intercal}\bfbeta$ or an unspecified nonparametric form. The subscript $\xi$ indicates that the operator is taken under the underlying prediction model, also called the outcome regression model. The error terms $\varepsilon_i$ are independent with $E_\xi(\varepsilon_i) = 0$ and $V_\xi(\varepsilon_i) = v(\bfx_i) \sigma^2$. The variance function $v(\bfx_i)$ has a known form, and the homogeneous variance structure with $v(\bfx_i)=1$ might be used for certain applications.  

Under the assumption {\bf A1}, we have $E_\xi(y_i \mid \bfx_i, R_i=1) = E_\xi(y_i \mid \bfx_i) = m(\bfx_i)$. The dataset $\{(\bfx_i,y_i), i\in \mS_{\A}\}$ from the non-probability sample can be used to build the model (\ref{reg}). For the linear regression model where $m(\bfx_i)=\bfx_i^{\intercal}\bfbeta$ and $v(\bfx_i)=1$, the least square estimator of $\bfbeta$ is given by 
\[
\hat\bfbeta  = \Big(\sum_{i=1}^N R_i \bfx_i \bfx_i^{\intercal}\Big)^{-1}\Big(\sum_{i=1}^NR_i\bfx_iy_i\Big) 
= \Big(\sum_{i\in \mS_{\A}} \bfx_i \bfx_i^{\intercal}\Big)^{-1}\Big(\sum_{i\in \mS_{\A}}\bfx_iy_i\Big) \,.
\]
The predicted value for $y_i$ with an associated $\bfx_i$ is given by $\hat{y}_i = \bfx_i^{\intercal}\hat\bfbeta$. The auxiliary information $\bfx_i$ from the reference probability sample $\mS_{\B}$, along with the survey weights $d_i^{\B}$, provides a regression prediction estimator for the population mean $\mu_y = N^{-1}\sum_{i=1}^Ny_i$, which is constructed as 
\[
\hat{\mu}_{\REG} = \frac{1}{{\hat{N}}^{\B}}\sum_{i\in \mS_{\B}} d_i^{\B} \hat{y}_i = \frac{1}{{\hat{N}}^{\B}}\sum_{i\in \mS_{\B}} d_i^{\B} \bfx_i^{\intercal}\hat\bfbeta\,,
\]
where $\hat{N}^{\B}=\sum_{i\in \mS_{\B}}d_i^{\B}$ is the estimated population size using the reference probability sample. The estimator $\hat{\mu}_{\REG}$ is approximately unbiased estimator for $\mu_y$ under both the regression model and the probability sampling design for $\mS_{\B}$. If $N$ is known and $\hat{N}^{\B}$ is replaced by $N$ in the construction of $\hat{\mu}_{\REG}$, the estimator is exactly unbiased. The estimator $\hat{\mu}_{\REG}$ with the estimated $N$ is the well-known H\'{a}jek estimator and is preferred to use in practice even if $N$ is known. The regression prediction estimator $\hat{\mu}_{\REG}$ tends to perform well if the model $y_i = \bfx_i^{\intercal}\bfbeta + \varepsilon_i$ has strong prediction power (Kang and Schafer, 2007). 

The regression prediction estimator is a special case of the so-called ``{\em mass imputation}'' methods. The probability sample $\mS_{\B}$ can be viewed as a sample with the response values $y_i$ missing for all units. The estimator $\hat{\mu}_{\REG}$ uses $\bfx_i^{\intercal}\hat\bfbeta$ as an imputed value for $y_i$. In general, an imputation based estimator can be constructed as 
$\hat{\mu}_{\SM} = ({\hat{N}}^{\B})^{-1}\sum_{i\in \mS_{\B}} d_i^{\B} y_i^*$, 
where $y_i^*$ is an imputed value for $y_i$, and the subscript ``SM'' indicates ``{\em sample matching}'' (Rivers 2007; Vavreck and Rivers 2008). A popular sample matching approach is the nearest neighbor imputation method. For each $i\in \mS_{\B}$, we let $y_i^*=y_j$, where $j\in \mS_{\A}$ and $\bfx_j$ minimizes the distance $\|\bfx_k -\bfx_i\|$ for all $k\in \mS_{\A}$. In other words, we match each missing $y_i$, $i\in \mS_{\B}$ with an observed $y_j$, $j\in \mS_{\A}$. 

Brick (2015) presented a compositional approach to non-probability samples with a focus on poststratification. He also discussed diagnostics and model checking for prediction approaches, which is also the most crucial aspect of sample matching methods.

\bigskip

\setcounter{section}{3}
\setcounter{equation}{0}
\no {\bf 3. DOUBLY ROBUST INFERENTIAL PROCEDURES}

\medskip

\no
In this section, we first present a rigorous procedure for the estimation of the propensity scores. We then discuss construction of doubly robust estimators of the finite population mean using the estimated propensity scores as well as an outcome regression model. 

\medskip

\no {\bf 3.1  Estimation of Propensity Scores}

\medskip

\no
Consider the hypothetical situation where $\bfx_i$ is observed for all units in the finite population $\mU$ while $y_i$ is only observed for the non-probability sample $\mS_{\A}$.  Estimation of the propensity scores under this scenario becomes the standard missing data problem with observations $\{(R_i, R_iy_i, \bfx_i), i=1,2,\cdots,N\}$. Suppose that the propensity scores can be modelled parametrically as 
$\pi_i^{\A}= P(R_i = 1 \mid \bfx_i) = \pi(\bfx_i,\bftheta_0)$, where $\bftheta_0$ is the true value of the unknown model parameters. The maximum likelihood estimator of $\pi_i$ is computed as $\hat{\pi}_i^{\A} = \pi(\bfx_i,\hat\bftheta)$, where $\hat{\bftheta}$ maximizes the log-likelihood function 
\begin{eqnarray}
l(\bftheta)&=&\sum_{i=1}^{N} \Big\{R_i\log{\pi_i^{\A}}+ (1-R_i )\log{(1-\pi_i^{\A})} \Big\} \nonumber\\
 &=& \sum_{i\in \mS_{\A}} \log\left\{\frac{\pi(\bfx_i,\bftheta)}{1-\pi(\bfx_i,\bftheta)}\right\}   
 + \sum_{i=1}^{N} \log\big\{1-\pi(\bfx_i,\bftheta)\big\}\,.
 \label{l1}
 \end{eqnarray} 
However, the log-likelihood function specified in (\ref{l1}) cannot be used in practice since we do not observe $\bfx_i$ for all units in the finite population. This is where we need the reference probability sample $\mS_{\B}$ with information on $\bfx$. Instead of using $l(\bftheta)$, we compute the estimator $\hat\bftheta$ by maximizing the following pseudo log-likelihood function
\begin{equation}
l^*(\bftheta)=\sum_{i\in \mS_{\A}} \log\left\{\frac{\pi(\bfx_i,\bftheta)}{1-\pi(\bfx_i,\bftheta)}\right\}  + 
\sum_{i \in \mS_{\B}} d_i^{\B}\log\big\{1-\pi(\bfx_i,\bftheta)\big\}\,,  
\label{l2}
\end{equation}
where the population total $\sum_{i=1}^{N} \log\{1-\pi(\bfx_i,\bftheta)\}$ in $l(\bftheta)$ is replaced by the Horvitz-Thompson estimator $\sum_{i \in \mS_{\B}} d_i^{\B}\log\{1-\pi(\bfx_i,\bftheta)\}$ using the reference sample $\mS_{\B}$. Note that $d_i^{\B} = 1/\pi_i^{\B}$, where $\pi_i^{\B} = P(i\in \mS_{\B})$, are the survey weights from the probability sample. 

Under a logistic regression model for the propensity scores where $\pi_i^{\A}={\pi(\bfx_i,\bftheta_0)}=\exp{(\bfx_i^{\intercal}\bftheta_0)}/\{1+ \exp{(\bfx_i^{\intercal}\bftheta_0)}\}$, the pseudo log-likelihood function (\ref{l2}) becomes 
\[
l^*(\bftheta) =  \sum_{i\in \mS_{\A}} \bfx_i^{\intercal}\bftheta -\sum_{i\in \mS_{\B}} d_i^{\B} \log{\big\{{1+ \exp{(\bfx_i^{\intercal}\bftheta)}}\big\}} \,.
\]
The maximum pseudo likelihood estimator $\hat\bftheta$ can be obtained by solving the score equations $U(\bftheta) = {\bf 0}$ where 
\[
U(\bftheta) = \frac{\partial}{\partial \bftheta} l^*(\bftheta)=\sum_{i\in \mS_{\A}}\bfx_i- \sum_{i \in \mS_{\B}} d_i^{\B} \pi(\bfx_i,\bftheta) \bfx_i \,.
\]
The solution can be found by using the following Newton-Raphson iterative procedure 
\[
\bftheta^{(m+1)} = \bftheta^{(m)} + \Big\{H(\bftheta^{(m)})\Big\}^{-1}U(\bftheta^{(m)})\,,
\]
where $H(\bftheta)=\sum_{i\in \mS_{\B}}d_i^{\B} \pi(\bfx_i,\bftheta)\{1- \pi(\bfx_i,\bftheta)\}\bfx_i\bfx_i^{\intercal}$ and the initial value for the iteration can be chosen as $\bftheta^{(0)}={\bf 0}$.

\medskip

\no {\bf 3.2  Inverse Probability Weighted Estimator}

\medskip

\no
The inverse probability weighted (IPW) estimator is the most successful adoption of the Horvitz-Thompson (HT) estimator for missing data problems and causal inferences. The HT estimator was originally proposed by Horvitz and Thompson (1952) for a finite population with probability survey samples where the weights are determined by the sampling design. The IPW estimator, however, requires modeling on the propensity scores and its use in the survey context is also referred to as the quasi-randomization approach (Kott, 1994).

The estimated propensity scores $\hat{\pi}_i^{\A} = \pi(\bfx_i,\hat\bftheta)$, $i \in \mS_{\A}$ can be used to compute two versions of the IPW estimator for the population mean $\mu_y$, depending on whether the population size $N$ is known or not:
\begin{equation}
\hat{\mu}_{\IPWone} = \frac{1}{N} \sum_{i \in \mS_{\A}} \frac{y_i}{{\hat{\pi}_i}^{\A}} \;\;\;\;\; {\rm and } \;\;\;\;\;
\hat{\mu}_{\IPWtwo} = \frac{1}{\hat{N}^{\A}}\sum_{i \in \mS_{\A}} \frac{y_i}{{\hat{\pi}_i}^{\A}}\,, 
\label{ipw}
\end{equation}
where $\hat{N}^{\A}=\sum_{i \in \mS_{\A}} (\hat{\pi}_i^{\A})^{-1}$. The estimator $\hat{\mu}_{\IPWtwo}$ is the so-called H\'{a}jek estimator in survey sampling and has certain advantages over the estimator $\hat{\mu}_{\IPWone}$ even if the population size $N$ is known. 

We consider the following asymptotic framework for theoretical development. Suppose that there is a sequence of finite populations ${\mU_{\nu}}$ of size $N_{\nu}$, indexed by $\nu$. Associated with each ${\mU_{\nu}}$ are a non-probability sample $\mS_{\A, \nu}$ of size $n_{\A, \nu}$ and a probability sample $\mS_{\B, \nu}$ of size $n_{\B, \nu}$. The population size $N_{\nu} \rightarrow \infty$ and the sample sizes $n_{\A, \nu} \rightarrow \infty$ and $n_{\B, \nu} \rightarrow \infty$ as $\nu \rightarrow \infty$. For notational simplicity the index $\nu$ is suppressed for the rest of the paper and the limiting process is represented by $N \rightarrow \infty$. The properties of the IPW estimators, summarized in the Theorem below, are developed under both the model for the propensity scores and the survey design for the probability sample $\mS_{\B}$. Proof of the theorem is given in the Appendix. 

\medskip

\noindent
{\bf Theorem 1} .\ {\em Under the regularity conditions {\bf A1}--{\bf A3} and {\bf C1}--{\bf C6} specified in the Appendix and assuming the logistic regression model for the propensity scores, we have $\hat{\mu}_{\IPWone} - \mu_y = O_p(n_{\A}^{-1/2})$, $\hat{\mu}_{\IPWtwo} - \mu_y = O_p(n_{\A}^{-1/2})$, $Var(\hat{\mu}_{\IPWone}) = V_{\IPWone} +o(n_{\A}^{-1})$,  
$Var(\hat{\mu}_{\IPWtwo}) = V_{\IPWtwo}  +o(n_{\A}^{-1})$, where 
\begin{equation}
V_{\IPWone} = \frac{1}{N^2}\sum_{i =1}^{N}(1-\pi_{i}^{\A})\pi_{i}^{\A}\bigg(\frac{y_i}{\pi_{i}^{\A}}-{\bf b}_1^{\intercal}\bfx_i\bigg)^2 +  {\bf b}_1^{\intercal} {\bf D}{\bf b}_1\,, 
\label{V1}
\end{equation}
\begin{equation}
V_{\IPWtwo} = \frac{1}{N^2}\sum_{i =1}^{N}(1-\pi_{i}^{\A})\pi_{i}^{\A}\bigg(\frac{y_i-\mu_y}{\pi_{i}^{\A}}-{\bf b}_2^{\intercal}\bfx_i\bigg)^2+{\bf b}_2^{\intercal} {\bf D}{\bf b}_2\,, \label{V2}
\end{equation}
where $\pi_{i}^{\A} = \pi(\bfx_i, \bftheta_0)$, 
${\bf b}_1^{\intercal}={\big\{N^{-1} \sum_{i =1}^{N}(1-\pi_{i}^{\A})y_i\bfx_i^{\intercal}\big\}}\big\{ N^{-1}\sum_{i =1}^{N}\pi_{i}^{\A}(1-\pi_{i}^{\A})\bfx_i\bfx_i^{\intercal}\big\}^{-1}$,
${\bf b}_2^{\intercal}={\big\{N^{-1}\sum_{i =1}^{N}(1-{\pi_{i}^{\A}})(y_i-\mu_y) \bfx_i^{\intercal}\big\}}\big\{N^{-1}\sum_{i =1}^{N}\pi_{i}^{\A}(1-\pi_{i}^{\A})\bfx_i\bfx_i^{\intercal}\big\}^{-1}$, \ 
and \ ${\bf D} = N^{-2} $ $V_p\big(\sum_{i \in  \mS_{\B}}d_i^{\B}\pi_{i}^{\A}\bfx_i \big)$, where $V_p(\cdot)$ denotes the design-based variance under the probability sampling design for $\mS_{\B}$. 
}

Under slightly tightened conditions for the propensity score model and the survey design on the sample $\mS_{\B}$ where both $N^{-1}\sum_{i=1}^NR_i\bfx_i$ and $N^{-1}\sum_{i\in \mS_{\B}}d_i^{\B}\pi_i^{\A}\bfx_i$ are asymptotically normally distributed, we have that both 
$(V_{\IPWone})^{-1/2}(\hat{\mu}_{\IPWone} - \mu_y)$ and $(V_{\IPWtwo})^{-1/2}(\hat{\mu}_{\IPWtwo} - \mu_y)$ converge to $N(0,1)$ in distribution.

\medskip

\no {\bf 3.3  Doubly Robust Estimator}

\medskip

\no
The IPW estimators are sensitive to misspecified models for the propensity scores, especially when certain units have very small values in $\hat{\pi}_i^{\A}$. See, for instance, Tan (2007) for further discussion. The efficiency and the robustness of IPW estimators can be improved by incorporating a prediction model for the response variable.  Robins et al. (1994) identified a class of augmented inverse probability weighted (AIPW) estimators under the two-model framework, and showed the improved efficiency of AIPW estimators over the IPW estimators when both models are correct. Scharfstein et al. (1999) further noticed that this class of  AIPW estimators remains consistent as long as one of the two models is correctly specified. This is the so-called double robustness property that is widely studied in the recent literature on missing data problems. 

Consider a parametric model $E_\xi(y \mid \bfx)=m(\bfx,\bfbeta_0)$ for the response $y$ given the $\bfx$, where the subscript $\xi$ indicates the model for the outcome regression. The general form of the doubly robust (DR) estimator for $\mu_y$ is given by 
\begin{equation}
\hat{\mu}_{\DR} = \frac{1}{N}\sum_{i=1}^{N}\frac{R_i\big\{y_i-m(\bfx_i, \hat{\bfbeta})\big\}}{\pi(\bfx_i, {\hat{\bftheta}})}+\frac{1}{N}\sum_{i=1}^{N}m(\bfx_i, \hat{\bfbeta})\,, 
\label{gdr}
\end{equation}
where $\hat{\bftheta}$ and  $\hat{\bfbeta}$  are consistent estimators of the true parameters $\bftheta_0$ and ${\bfbeta_0}$ under each of the two models. The estimator $\hat{\mu}_{\DR}$ given by (\ref{gdr}) is identical to the model-assisted ``generalized difference estimator'' discussed in Wu and Sitter (2001) under scenarios where the complete auxiliary information $\{\bfx_1,\cdots,\bfx_{\N}\}$ is available. Our proposed doubly robust estimator for $\mu_y$ under the current setting is given by 
\begin{equation}
\hat{\mu}_{\DRone}= \frac{1}{N}\sum_{i\in \mS_{\A}} d_i^{\A} \big\{y_i-m(\bfx_i, \hat{\bfbeta})\big\}+\frac{1}{N}\sum_{i \in \mS_{\B}}d_i^{\B}m(\bfx_i, \hat{\bfbeta})\, ,
\label{dr1}
\end{equation}
where $d_i^{\A} = 1/\pi(\bfx_i, {\hat{\bftheta}})$. An alternative estimator using the estimated population size is given by 
\begin{equation}
\hat{\mu}_{\DRtwo}= \frac{1}{\hat{N}^{\A}}\sum_{i\in \mS_{\A}} d_i^{\A} \big\{y_i-m(\bfx_i, \hat{\bfbeta})\big\}+\frac{1}{\hat{N}^{\B}}\sum_{i \in \mS_{\B}}d_i^{\B}m(\bfx_i, \hat{\bfbeta})\, ,
\label{dr2}
\end{equation}
where $\hat{N}^{\A} = \sum_{i\in \mS_{\A}}d_i^{\A}$ and $\hat{N}^{\B} = \sum_{i\in \mS_{\B}}d_i^{\B}$.

The development of theoretical properties of $\hat{\mu}_{\DRone}$ and $\hat{\mu}_{\DRtwo}$ requires a joint randomization framework involving the propensity score model for $\mS_{\A}$, the outcome regression model $\xi$ and the probability sampling design for $\mS_{\B}$. However, an important feature of the estimators $\hat{\mu}_{\DRone}$ and $\hat{\mu}_{\DRtwo}$ is that the estimator $\hat{\bfbeta}$ does not have any impact on the asymptotic variance regardless whether the regression model $\xi$ is correctly specified or not. This feature allows us to practically drop the regression model in developing the asymptotic variance.   
We  assume that $\hat{\bfbeta}=\bfbeta^{*}+O_p(n_{\A}^{-1/2})$ for some fixed $\bfbeta^*$. The value of $\bfbeta^*$ is the same as the true parameter ${\bfbeta_0}$ when the regression model is correctly specified but has no practical meanings otherwise. We consider the logistic regression model for the propensity scores and focus on the practically useful estimator $\hat{\mu}_{\DRtwo}$ in the following theorem. 
Let $\pi_i^{\A} = \pi(\bfx_i,\bftheta_0)$ be the true value of the propensity score. 

\medskip

\noindent
{\bf Theorem 2.} \ {\em The estimator $\hat{\mu}_{\DRtwo}$ is doubly robust in the sense that it is a consistent estimator of $\mu_y$ if either the propensity score model or the outcome regression model is correctly specified. Furthermore, under the regularity conditions {\bf C1}--{\bf C6} specified in the Appendix and the correctly specified logistic regression model for the propensity scores, we have $Var\big(\hat{\mu}_{\DRtwo}\big) = V_{\DRtwo} + o(n_{\A}^{-1})$ where 
\begin{equation}
V_{\DRtwo} = \frac{1}{N^2}\sum_{i =1}^{N}(1-\pi_{i}^{\A})\pi_{i}^{\A} \big[\{y_i-m(\bfx_i,\bfbeta^{*})-h_{\N}\}/\pi_{i}^{\A} -{\bf b}_3^{\intercal}\bfx_i\big]^2 + W\,, 
\label{aspd2}
\end{equation}
where 
${\bf b}_3^{\intercal}=\big[N^{-1}\sum_{i =1}^{N}(1-\pi_{i}^{\A})\big\{y_i -m(\bfx_i,\bfbeta^{*})-h_{\N}\big\}\bfx_i^{\intercal}\big]
\big\{N^{-1}\sum_{i =1}^{N}\pi_{i}^{\A}(1-\pi_{i}^{\A})\bfx_i\bfx_i^{\intercal}\big\}^{-1}$, 
$h_{\N} = N^{-1}\sum_{i =1}^{N}\big\{ y_i -m(\bfx_i,\bfbeta^{*})\big\}$, 
and $W = N^{-2}V_p\big(\sum_{i\in \mS_{\B}}d_i^{\B}t_i \big)$ is the design-based variance with 
$t_i=\pi_{i}^{\A}\bfx_i^{\intercal} {\bf b}_3 + m(\bfx_i,\bfbeta^{*}) -N^{-1}\sum_{i =1}^{N}m(\bfx_i,\bfbeta^{*})$.
}

\medskip
 
Efficiency comparisons between the IPW estimators and the DR estimators is not a straightforward topic and has been studied extensively in the missing data literature. See, for instance, Robins et al. (1994), Tan (2007), Cao et al. (2009), among others. The doubly robust estimator is constructed through the residual variable $e_i = y_i - m(\bfx_i,\bfbeta)$ and usually has smaller variance if the regression model provides a good fit to the non-probability survey data. 

\bigskip

\setcounter{section}{4}
\setcounter{equation}{0}
\no {\bf 4. VARIANCE ESTIMATION}

\medskip

\no
The asymptotic variance formulas presented in Section 3 provide a simple plug-in method for variance estimation. However, the asymptotic variance formulas for the doubly robust estimators are derived under the assumed model for the propensity scores. The plug-in variance estimator becomes inconsistent when the outcome regression model is correctly specified but the propensity score model is misspecified. The doubly robust variance estimation technique proposed by Kim and Haziza (2014) is a preferred approach and can be implemented under the current context. 

\medskip

\no {\bf 4.1  Plug-in Variance Estimators}

\medskip

\no
We show the details of the plug-in variance estimator for the IPW estimator $\hat{\mu}_{\IPWtwo}$. Using the asymptotic variance formula (\ref{V2}) presented in Theorem 1, the first piece 
$ N^{-2}\sum_{i =1}^{N}(1-\pi_{i}^{\A})\pi_{i}^{\A}\big\{(y_i-\mu_y)/\pi_{i}^{\A}-{\bf b}_2^{\intercal}\bfx_i\big\}^2$ 
can be consistently estimated by 
\[
\frac{1}{N^2}\sum_{i \in \mS_{\A}}(1-\hat{\pi}_i^{\A}){\Big(\frac{y_i - \hat{\mu}_{\IPWtwo}}{\hat{\pi}_i^{\A}}-\hat{\bf b}_2^{\intercal}\bfx_i\Big)}^2\,,
\]
where $N$ might be replaced by $\hat{N}^{\A}$ if necessary, and 
\[
\hat{\bf b}_2^{\intercal} = {\bigg\{\sum_{i \in \mS_{\A}}\Big(\frac{1}{{\hat{\pi}_i}^{\A}}-1\Big)(y_i-\hat{\mu}_{\IPWtwo})\bfx_i^{\intercal}\bigg\}}\bigg\{\sum_{i \in \mS_{\B}}d_i^{\B}\hat{\pi}_i^{\A}(1-\hat{\pi}_i^{\A})\bfx_i\bfx_i^{\intercal}\bigg\}^{-1}\,.
\]
The second piece ${\bf b}_2^{\intercal} {\bf D}{\bf b}_2$ can be estimated by $\hat{\bf b}_2^{\intercal} \hat{\bf D} \hat{\bf b}_2$, where $\hat{\bf D}$ is the design-based variance estimator and is given by 
\[
\hat{\bf D} = \frac{1}{N^2}\sum_{i \in \mS_{\B}}\sum_{j \in \mS_{\B}} \frac{\pi_{ij}^{\B}-\pi_i^{\B} \pi_j^{\B}}{\pi_{ij}^{\B}}\, \frac{\hat{\pi}_i^{\A}}{\pi_i^{\B}} \, \frac{\hat{\pi}_{j}^{\A}}{\pi_j^{\B}} \bfx_i \bfx_{j}^{\intercal} \,,
\]
where $\pi_i^{\B}$ and $\pi_{ij}^{\B}$ are the first and second order inclusion probabilities for the probability sample $\mS_{\B}$. For certain sampling designs, the second order inclusion probabilities $\pi_{ij}^{\B}$ might not be needed and approximate estimators for the design-based variance 
$V_p\big(\sum_{i\in \mS_{\B}} d_i^{\B}\pi_i^{\A}\bfx_i\big)$ are available from the survey sampling literature. 

When the propensity score model is valid, a plug-in variance estimator for the doubly robust estimator $\hat{\mu}_{\DRtwo}$ can be similarly constructed based on the asymptotic variance formula presented in Theorem 2. 

\medskip

\no {\bf 4.2 Doubly Robust Variance Estimator}

\medskip

\no
Let $E_q$, $E_{\xi}$, $E_p$, $V_q$, $V_{\xi}$ and $V_p$ denote the expectation and variance under the propensity score model $q$, the outcome regression model $\xi$ and the probability sampling design $p$ for $\mS_{\B}$, respectively. We have $E_q(R_i \mid \bfx_i) = \pi(\bfx_i,\bftheta)$ and 
$V_q(R_i \mid \bfx_i) = \pi(\bfx_i,\bftheta)\{1-\pi(\bfx_i,\bftheta)\}$. We also have $E_{\xi}(y_i \mid \bfx_i) = m(\bfx_i,\bfbeta)$ and 
$V_{\xi}(y_i \mid \bfx_i) = v(\bfx_i) \sigma^2$.

The uncertainty of not knowing which of the two models $q$ and $\xi$ is valid for doubly robust estimators presents a real challenge for variance estimation. 
The concept of doubly robust variance estimation is appealing and has been discussed by several authors, including Haziza and Rao (2006) and Kim and Park (2006). The variance estimator is doubly robust if it is approximately unbiased for the variance of the doubly robust point estimator when one of the models $q$ or $\xi$ is correctly specified. In this section, we illustrate how to implement the method proposed by Kim and Haziza (2014) under the current setting on non-probability survey samples. Let the doubly robust point estimator be computed as 
\begin{equation}
\hat{\mu}_{\KH}= \frac{1}{N}\sum_{i=1}^{N}\frac{R_i\big\{y_i-m(\bfx_i, \hat{\bfbeta})\big\}}{\pi(\bfx_i, {\hat\bftheta})}+\frac{1}{N}\sum_{i \in \mS_{\B}}d_i^{\B}m(\bfx_i, \hat{\bfbeta}) \,,
\label{KH}
\end{equation}
where the subscript ``KH'' indicates ``Kim-Haziza''. 
The form of the estimator is identical to $\hat{\mu}_{\DRone}$ given in (\ref{dr1}). However, instead of estimating $\bftheta$ and $\bfbeta$ separately using the propensity score model and the regression model, we estimate $(\bftheta, \bfbeta)$ by $(\hat\bftheta, \hat\bfbeta)$ by solving the following system of estimating equations:
\begin{eqnarray}
\frac{1}{N} \sum_{i =1}^{N}R_i\bigg(\frac{1}{\pi(\bfx_i,\bftheta)}-1\bigg)\big\{y_i-m(\bfx_i,\bfbeta)\big\}\bfx_i &=& {\bf 0} \,,
\label{kheq1}\\
\frac{1}{N} \sum_{i =1}^{N}\frac{R_i}{{\pi}(\bfx_i,\bftheta)}\dot{m}(\bfx_i,\bfbeta)-\frac{1}{N}\sum_{i \in \mS_{\B}}d_i^{\B}\dot{m}(\bfx_i,\bfbeta) &=& {\bf 0} \,,
\label{kheq2}
\end{eqnarray}
where $\dot{m}(\bfx,\bfbeta) = \partial m(\bfx,\bfbeta)/\partial \bfbeta$. 
It is assumed that the outcome regression model contains an intercept and the first component of $\bfx$ is 1, and $\pi(\bfx_i,\bftheta) = \exp{(\bfx_i^{\intercal}\bftheta)}/\{1+ \exp{(\bfx_i^{\intercal}\bftheta)}\}$. Under the linear regression model where $m(\bfx_i,\bfbeta) = \bfx_i^{\intercal}\bfbeta$, we have $\dot{m}(\bfx_i,\bfbeta)=\bfx_i$ and the equation system from  (\ref{kheq2}) becomes 
\[
\frac{1}{N} \sum_{i =1}^{N}\frac{R_i\bfx_i}{{\pi}(\bfx_i,\bftheta)} - \frac{1}{N}\sum_{i \in \mS_{\B}}d_i^{\B} \bfx_i = {\bf 0} \,,
\]
which is slightly different from the score equations $U(\bftheta) = {\bf 0}$ presented in Section 3.1 under the logistic regression model.

There are two major consequences from this approach. The first is the asymptotic expansion to $\hat{\mu}_{\KH}$ given by 
\begin{eqnarray}
\hat{\mu}_{\KH} - \mu_y &=& \frac{1}{N}\sum_{i =1}^{N}\Big( \frac{R_i}{\pi_{i}^*}-1\Big)\Big\{y_i-m(\bfx_i,\bfbeta^{*})\Big\}+ \frac{1}{{N}}\sum_{i \in \mS_{\B}}d_i^{\B}m(\bfx_i,\bfbeta^{*})\nonumber\\
& & -\frac{1}{N}\sum_{i =1}^{N}m(\bfx_i,\bfbeta^{*})+o_p(n_{\A}^{-1/2})\,, \label{vkh}
\end{eqnarray}
where $\bfbeta^{*}$ and $\bftheta^{*}$ are the limit of  $\hat{\bfbeta}$ and $\hat{\bftheta}$, respectively,  and $\pi_{i}^*=\pi(\bfx_i, \bftheta^{*})$. The second consequence is the construction of a variance estimator which is approximately unbiased under the joint randomization involving either $q$ or $\xi$ (but not both) and the sampling design $p$, as shown below. 

We first derive a variance estimator for $\hat{\mu}_{\KH}$ under the joint randomization $q$ and $p$. It follows from (\ref{vkh}) that 
$V_{qp}(\hat{\mu}_{\KH}) = W_1 + W_2 + o(n_{\A}^{-1})$ where 
\[
W_1 = V_q\Bigg[\frac{1}{N}\sum_{i =1}^{N}\Big( \frac{R_i}{\pi_{i}^*}-1\Big)\Big\{y_i-m(\bfx_i,\bfbeta^{*})\Big\}\Bigg] 
=\frac{1}{N^2}\sum_{i=1}^N \frac{(1-\pi_i^{\A})\pi_i^{\A}}{(\pi_i^*)^2}\Big\{y_i-m(\bfx_i,\bfbeta^{*})\Big\}^2\,,
\]
and $W_2=V_p\big\{N^{-1}\sum_{i \in \mS_{\B}}d_i^{\B}m(\bfx_i,\bfbeta^{*})\big\}$. The second term $W_2$ is the design-based variance and can be estimated using standard methods for the sample $\mS_{\B}$. Let $\hat{W}_2$ be the estimator for $W_2$. We can estimate the first term $W_1$ by 
\begin{equation}
\hat{W}_1 = \frac{1}{N^2}\sum_{i =1}^{N}\frac{R_i(1-\hat{\pi}_i^{\A})}{(\hat{\pi}_i^{\A})^{2}}\big\{y_i-m(\bfx_i, \hat{\bfbeta})\big\}^{2} \,. 
\end{equation}

The asymptotic variance of $\hat{\mu}_{\KH}$ under the joint randomization $\xi$ and $p$ is given by 
$V_{\xi p}(\hat{\mu}_{\KH} - \mu_y) = K_1 + W_2 + o(n_{\A}^{-1})$, where $W_2$ is the same design-based variance defined in $V_{qp}(\hat{\mu}_{\KH})$ and 
\[
K_1 = V_\xi\bigg[\frac{1}{N}\sum_{i =1}^{N}\Big(\frac{R_i}{\pi_{i}^*}-1\Big)\Big\{y_i-m(\bfx_i,\bfbeta^{*})\Big\} \bigg]  
= \frac{1}{N^2}\sum_{i=1}^N \Big(\frac{R_i}{\pi_{i}^*}-1\Big)^2 \sigma_i^2\,,
\]
where $ \sigma_i^2 = V_{\xi}(y_i \mid \bfx_i) = v(\bfx_i)\sigma^2$. It is apparent that $\hat{W}_1$ is not a valid estimator for $K_1$ under the model $\xi$, and the bias is given by 
\[
E_{\xi}\big(\hat{W}_1\big) - K_1 = \frac{1}{N^2}\sum_{i=1}^N \Big(\frac{R_i}{\pi_{i}^*}-1\Big)\sigma^2_i + o(n_{\A}^{-1})\,.
\]
An important observation is that the bias is non-negligible under the outcome regression model $\xi$ but the expectation of the leading term in the bias under the propensity score model $q$ is approximately zero. This leads to the following doubly robust variance estimator for $\hat{\mu}_{\KH}$:
\[
v(\hat{\mu}_{\KH}) = \hat{W}_1+\hat{W}_2 - \frac{1}{N^2}\bigg\{\sum_{i\in \mS_{\A}}\frac{\hat{\sigma}_i^2}{\hat{\pi}_i^{\A}}-\sum_{i\in \mS_{\B}}d_i^{\B}\hat{\sigma}_i^2 \bigg\} \,.
\]

\bigskip

\setcounter{section}{5}
\setcounter{equation}{0}
\no {\bf 5. SIMULATION STUDIES}

\medskip

\no
We consider finite populations of size $N=20,000$ with the response variable $y$ and auxiliary variables $\bfx$ following the regression model ($\xi$)
\[
y_i=2+ x_{1i}+x_{2i}+x_{3i}+x_{4i}+\sigma \varepsilon_i \,, \;\;\; i=1,2,\cdots,N\,,
\]
where $x_{1i}=z_{1i}$, 
$x_{2i}=z_{2i}+0.3x_{1i}$,
$x_{3i}=z_{3i}+0.2(x_{1i}+x_{2i})$,
$x_{4i}=z_{4i}+0.1(x_{1i}+x_{2i}+x_{3i})$, with 
$z_{1i} \sim Bernoulli(0.5)$, 
$z_{2i} \sim Uniform(0,2)$, 
$z_{3i} \sim Exponential(1)$ and 
$z_{4i} \sim \chi^2(4)$. 
The error term $\varepsilon_i$'s are independent and identically distributed (iid) as $N(0,1)$, with values of $\sigma$ chosen such that the correlation coefficient $\rho$ between $y$ and the linear predictor 
$\bfx^{\intercal} \bfbeta$ is controlled at 0.3, 0.5 and 0.8 for the simulation studies. The parameter of interest is the finite population mean $\mu_y$. 

The true propensity scores $\pi_i^{\A}$ for the non-probability sample $\mS_{\A}$ follow the logistic regression model ($q$)
\[
\log\bigg(\frac{\pi_i^{\A}}{1-\pi_i^{\A}}\bigg) = \theta_0+ 0.1x_{1i}+0.2x_{2i}+0.1x_{3i}+0.2x_{4i}\,,
\]
where $\theta_0$ is chosen such that $\sum_{i=1}^N\pi_i^{\A} = n_{\A}$ with the given target sample size $n_{\A}$. The non-probability sample $\mS_{\A}$ is selected by the Poisson sampling method with inclusion probabilities specified by $\pi_i^{\A}$ and the target sample size $n_{\A}$. 

The probability sample $\mS_{\B}$ with the target size $n_{\B}$ is also taken by the Poisson sampling method with the inclusion probabilities $\pi_i^{\B}$ proportional to $z_i = c + x_{3i} + 0.03y_i$. The value of $c$ is chosen to control the variation of the survey weights such that $\max z_i / \min z_i = 50$. 

We consider three scenarios for model specifications: (i) Both models $\xi$ and $q$ are correctly specified, denoted as ``TT''; (ii) The outcome regression mode $\xi$ is misspecified but the propensity score model $q$ is correctly specified, denoted as ``FT''. The working model for $\xi$ is chosen as 
$m(\bfx_i, \bfbeta)=\beta_0+ \beta_1x_{1i}+\beta_2x_{2i}+\beta_3x_{3i}$, with $x_{4i}$ omitted from the model; (iii) The outcome regression model $\xi$ is correctly specified but the propensity score model $q$ is misspecified, denoted as ``TF''. The working model for $q$ is given by 
$\log\{\pi_i^{\A}/(1-\pi_i^{\A}\} = \theta_0 + \theta_1x_{1i}+ \theta_2x_{2i}+ \theta_3x_{3i}$, with $x_{4i}$ omitted from the model. We consider four different sample size combinations for $(n_{\A},n_{\B})$ with $n_{\A}$ and $n_{\B}$ equaling either 500 or 1000. 

The first simulation study evaluates the performance of various point estimators for $\mu_y$. We focus on $\hat{\mu}_{\IPWone}$,  $\hat{\mu}_{\IPWtwo}$,   $\hat{\mu}_{\REG}$ and $\hat{\mu}_{\DRtwo}$ discussed in the paper. We also include three other estimators for the purpose of comparisons: $\hat{\mu}_{\A} = \bar{y}$, which is the simple sample mean from $\mS_{\A}$, and $\hat{\mu}_{\Cone}$ and $\hat{\mu}_{\Ctwo}$, which are computed in the same way as 
$\hat{\mu}_{\IPWone}$ and $\hat{\mu}_{\IPWtwo}$ but the propensity scores are estimated based on $\tilde{R}_i = 1$ if $i \in \mS_{\A}$ and $\tilde{R}_i=0$ if $i\in \mS_{\B}$. For a given estimator $\hat{\mu}$, its performance is evaluated through the relative bias (in percentage, \%RB) and mean squared error (MSE) computed as 
\[
\%RB=\frac{1}{B}\sum_{b=1}^{B}\frac{\hat{\mu}^{(b)}-\mu_y}{\mu_y}\times 100 \,, \;\;\;\;\;\;
MSE=\frac{1}{B}\sum_{b=1}^{B}{(\hat{\mu}^{(b)}-\mu_y)}^2 \,,
\]
where $\hat{\mu}^{(b)}$ is the estimator computed from the $b$th simulated sample and $B=10,000$ is the total number of simulation runs. 

Simulation results for $n_{\A}=500$ and $n_{\B}=1000$ are reported in Table \ref{tab1}. Results for other combinations of $(n_{\A},n_{\B})$ demonstrated similar patterns and are not included to save space. Major observations from Table \ref{tab1} can be summarized as follows. (1) The IPW estimator performs well under the correctly specified model $q$ for the propensity scores (``TT'' and ``FT''), with $\hat{\mu}_{\IPWtwo}$ having smaller MSE for all cases. Both estimators $\hat{\mu}_{\IPWone}$ and $\hat{\mu}_{\IPWtwo}$  collapse under a misspecified $q$ model. (2) The prediction estimator $\hat{\mu}_{\REG}$ performs very well under the correctly specified model $\xi$ for the outcome regression (``TT'' and ``TF'') but fails otherwise. (3) The doubly robust estimator $\hat{\mu}_{\DRtwo}$ has excellent performance under all three scenarios of model specifications. (4) The naive estimator $\hat{\mu}_{\A} = \bar{y}$ and the estimators $\hat{\mu}_{\Cone}$ and $\hat{\mu}_{\Ctwo}$ using the simple pooling method to estimate $\tilde{\pi}_i^{\A}$ do not provide valid results under any of the settings considered in the simulation. 

The second simulation study examines the behaviour of variance estimators. We consider variance estimators $v_{\IPWone}$ and $v_{\IPWtwo}$ associated with $\hat{\mu}_{\IPWone}$ and $\hat{\mu}_{\IPWtwo}$ based on Theorem 1 and the plug-in method described in Section 4.1. We also consider the variance estimator $v_{\PLUG}$ for $\hat{\mu}_{\DRtwo}$ using the plug-in method from Section 4.1 and the doubly robust variance estimator $v_{\KH}$ along with $\hat{\mu}_{\KH}$ discussed in Sections 4.2. The performance of a variance estimator $v$  along with the point estimator $\hat{\mu}$ is assessed by the percentage relative bias ($\%$RB) and the coverage probability ($\%$CP) computed as 
\[
\%RB=\frac{1}{B}\sum_{b=1}^{B}\frac{v^{(b)}-V}{V}\times100\,, \;\;\;\;\;\;
\%CP=\frac{1}{B}\sum_{b=1}^{B}I\big(\mu_y \in CI^{(b)}\big)\times 100\,,
\]
where $v^{(b)}$ is the variance estimator computed from the $b$th simulation sample, $V$ is the true variance of the point estimator obtained through a separate set of $B$ simulation runs, $I(\cdot)$ is the indicator function, and 
$CI^{(b)}=\big(\hat{\mu}^{(b)} -1.96(v^{(b)})^{1/2}, \hat{\mu}^{(b)} +1.96(v^{(b)})^{1/2}\big)$ is the $95\%$ confidence interval for $\mu_y$ based on the normal approximation. 

Simulation results for $n_{\A}=500$ and $n_{\B}=1000$ are reported in Table \ref{tab2}. The most important observation is that all variance estimator and the associated confidence intervals have excellent performance when the propensity score model is correctly specified (scenarios ``TT'' and ``FT''). The biases of the variance estimators are all small and the coverage probabilities of the $95\%$ confidence interval are close to the nominal value. When the propensity score model is misspecified (scenario ``TF''), the IPW point estimators are invalid and the related confidence intervals cannot be used. The plug-in variance estimator $v_{\PLUG}$ for $\hat{\mu}_{\DRtwo}$ has non-negligible negative bias, resulting in under-coverage for the confidence interval. The doubly robust variance estimator $v_{\KH}$ coupled with $\hat{\mu}_{\KH}$ has improvement over $v_{\PLUG}$, especially when the correlation between $y$ and $\bfx$ is strong (i.e., $\rho=0.80$).

\medskip

\setcounter{section}{6}
\setcounter{equation}{0}
\no {\bf 6. AN APPLICATION TO THE PEW SURVEY DATA}

\medskip

\no
In this section, we apply our proposed method to a dataset collected by the Pew Research Centre (http://www.pewresearch.org) in 2015. The dataset consists of nine non-probability samples with a total of 9,301 individuals and a wide range of measurements over 56 variables. The nine non-probability samples are supplied by eight vendors, which have different but unknown strategies in panel recruitment, sampling, incentives for participation, etc. In this analysis we treat the dataset as a single non-probability sample with $n_{\A}=9,301$. The dataset is referred to as PRC.

We consider two reference datasets as sources of auxiliary information. One is the Behavioral Risk Factor Surveillance System (BRFSS) survey dataset from 2015,  and the other is the Volunteer Supplement survey dataset from the Current Population Survey (CPS) in 2015. The BRFSS dataset (https://www.cdc.gov/brfss/index.html) contains 441,456 cases and a rich set of common variables with the PRC dataset (see Table \ref{tab3}). The CPS Volunteer Supplement dataset contains 80,075 cases and measurements on volunteering tendency, which are highly relevant to the response variables considered in the PRC dataset. The CPS dataset is often viewed as a reliable source of official statistics and contains a different set of common variables with the PRC dataset. 

We first examine the marginal distributions of some common variables listed in Table \ref{tab3} from the three datasets. Let $\hat{X}_{\PRC}$ denote the simple unadjusted sample mean (in $\%$) of a variable from the PRC sample; let $\hat{X}_{\BRFSS}$ and $\hat{X}_{\CPS}$ be the survey weighted estimates for the population mean from the BRFSS and the CPS datasets, respectively. While the two reference probability samples provide similar results over most of the variables, there is a clear discrepancy between the non-probability PRC sample and the two reference samples on age, race, origin and socioeconomic status. For instance, the PRC sample has $9.27\%$ participants with Hispanic/Latino origin and close to $42\%$ with a bachelor's degree or above, the corresponding numbers from the CPS sample are $15.60\%$ and $30.90\%$. 

We apply the proposed methods to estimate the population mean of seven response variables ($y$) listed in Tables \ref{tab4} and \ref{tab5}. The first six are binary variables and the last is a continuous variable. For the outcome regression model $\xi$ on $y$ given $\bfx$, we use the logistic regression model for the binary response and the linear regression model for the continuous response. The values of the estimators $\hat{\mu}_{\Cone}$, $\hat{\mu}_{\Ctwo}$, $\hat{\mu}_{\IPWone}$, $\hat{\mu}_{\IPWtwo}$, $\hat{\mu}_{\REG}$ and $\hat{\mu}_{\DRtwo}$ are computed for each $\mu_y$, and the computation is done separately with either the BRFSS or the CPS as the reference probability sample. The unadjusted simple sample mean $\hat{\mu}_{\A}$ from the non-probability PRC sample is also included. The survey weights for both BRFSS and CPS were calibrated by the data file producers such that $\hat{N}^{\B} = N$. 

Results from using a smaller common set of covariates which are available in all three datasets are presented in Table \ref{tab4}. The covariates are listed in the top part of Table \ref{tab3}, including Age, Gender, Race, Origin, Region, Marital Status, Employment and Education (High school or less, Bachelor's degree and above). The variable ``Age'' is treated as a continuous variable. The estimates from $\hat{\mu}_{\IPWtwo}$, $\hat{\mu}_{\REG}$ and $\hat{\mu}_{\DRtwo}$ (the last three columns) are very close to each other, indicating reasonable fit of models and the relevance of the auxiliary variables. The estimates are all different from the naive sample mean $\hat{\mu}_{\A}$. 
The estimator $\hat{\mu}_{\Cone}$ fails completely, shown the unreliability of using $N$ from other sources. The estimator $\hat{\mu}_{\Ctwo}$ also performs differently for most response variables. Another observation is that the results from using the two reference survey samples BRFSS or CPS are similar for most cases. 

We further include other available common covariates from each reference probability sample in addition to the set of common variables from all three datasets. Those additional variables are listed in the bottom part of Table \ref{tab3} for each of the reference samples. Results are computed separately from using BRFSS or CPS and are presented in Table \ref{tab5}. We see dramatic changes of results from using the two reference surveys. The estimates generally become smaller for using the CPS sample, and the estimates becomes significantly smaller for the two responses on volunteering: participation in school groups and in service organizations. This is probably due to the additional covariate on ``Volunteer Works'' from the CPS sample and the results should be more reliable under this setting.

\medskip

\setcounter{section}{7}
\setcounter{equation}{0}
\no {\bf 7. ADDITIONAL  REMARKS}

\medskip

\no
Assumptions {\bf A1}-{\bf A3} listed in Section 2.1 are part of the foundation for the estimation procedures presented in the paper. In practice, it is often difficult to decide whether the auxiliary variables $\bfx$ contain all the components for characterizing the selection mechanism. One of the general principles for collecting data using any non-probability method is to include essential auxiliary variables such as gender, age and measurements on socioeconomic status and other variables which not only provide tendencies for participation in the non-probability sample but also have the potential to be useful predictors for the response variables. The proposed estimation procedure calls for the availability of high quality probability survey samples with relevant auxiliary information. Census data and large scale probability samples collected by statistical agencies can serve as a rich source of information for statistical analysis of non-probability samples. As more and more data are collected by non-probability methods, the traditional survey-centric approach by many statistical agencies needs to evolve to stay relevant and effective for the new data era. 

The scenario of having zero propensity scores for certain units in the target population requires a careful evaluation of the population represented by the non-probability sample. This is the same issue of the under-coverage problem in probability sampling and the severity of the problem depends on the proportion of the uncovered  population units and the discrepancies between the two parts of the population in terms of the response variables. Corrections for biases due to under-coverage problems require additional source of information on the uncovered units.

\medskip

\setcounter{equation}{0}
\def\theequation{A.\arabic{equation}}

\no {\bf APPENDIX}

\medskip

\no
\textbf{A.1 \ Regularity Conditions}

\medskip

\no
Let $m(\bfx,\bfbeta)$ be the mean function of the outcome regression model. Let $\bfbeta_0$ be the true values of the model parameters. 

\begin{itemize}
\item[{\bf C1}] \  The population size $N$ and the sample sizes $n_{\A}$ and $n_{\B}$ satisfy $\lim_{N\rightarrow \infty} n_{\A}/N = f_{\A} \in (0,1)$ and $\lim_{N\rightarrow \infty} n_{\B}/N = f_{\B} \in (0,1)$. 
\item[{\bf C2}] \  For each $\bfx$, $\partial m(\bfx,\bfbeta)/\partial \bfbeta$ is continuous in $\bfbeta$ and $|\partial m(\bfx,\bfbeta)/\partial \bfbeta| \leq h(\bfx,\bfbeta)$ for $\bfbeta$ in the neighborhood of $\bfbeta_0$, and $N^{-1}\sum_{i=1}^Nh(\bfx_i,\bfbeta_0) = O(1)$. 
\item[{\bf C3}] \ For each $\bfx$, $\partial^2 m(\bfx,\bfbeta)/\partial \bfbeta \partial \bfbeta^{\intercal}$ is continuous in $\bfbeta$ and 
$\max_{j,l}|\partial^2 m(\bfx,\bfbeta)/\partial \beta_j\partial \beta_l | \leq k(\bfx,\bfbeta)$ for $\bfbeta$ in the neighborhood of $\bfbeta_0$, and $N^{-1}\sum_{i=1}^Nk(\bfx_i,\bfbeta_0) = O(1)$. 
\item[{\bf C4}] \ The finite population and the sampling design for $\mS_{\B}$ satisfy $N^{-1}\sum_{i\in \mS_{\B}}d_i^{\B}\bfu_i  - N^{-1}\sum_{i=1}^N \bfu_i = O_p(n_{\B}^{-1/2})$ for $\bfu_i = \bfx_i$, $y_i$ and $m(\bfx_i,\bfbeta)$. 
\item[{\bf C5}] \ There exist $c_1$ and $c_2$ such that $0<c_1\leq N\pi_i^{\A}/n_{\A}\leq c_2$ and $0<c_1\leq N\pi_i^{\B}/n_{\B}\leq c_2$ for all units $i$. 
\item[{\bf C6}] \ The finite population and the propensity scores satisfy $N^{-1}\sum_{i=1}^Ny_i^2=O(1)$, $N^{-1}\sum_{i=1}^N \|\bfx_i\|^3 = O(1)$, and 
$N^{-1}\sum_{i=1}^N \pi_i^{\A}(1-\pi_i^{\A})\bfx_i\bfx_i^{\intercal}$ is a positive definite matrix. 
\end{itemize}

Conditions {\bf C1} and {\bf C4} are commonly used in practice. Under condition {\bf C1},  we do not need to distinguish among $O_p(n_{\A}^{-1/2})$, $O_p(n_{\B}^{-1/2})$ and $O_p(N^{-1/2})$. Conditions {\bf C2} and {\bf C3} are the usual smoothness and boundedness conditions (Wu and Sitter, 2001). Condition {\bf C5} states that the inclusion probabilities for the samples $\mS_{\A}$ and $\mS_{\B}$ do not differ in terms of order of magnitude from simple random sampling. Condition {\bf C6} is the typical finite moment conditions and is used for making valid Taylor series expansions.  

\medskip

\no 
{\bf   A.2 \ Proof of Theorem 1}

\medskip

\no
Let ${\mathbf{\bfeta}}^{\intercal}=({\mu},{\bftheta}^{\intercal})$. The IPW estimators given in (\ref{ipw}) along with $\hat\bftheta$ for the propensity score model can be combined as $\hat{\mathbf{\bfeta}}^{\intercal}=(\hat{\mu},\hat{\bftheta}^{\intercal})$ which is the solution to the combined estimating equation system given by 
\begin{equation}
\label{eq}
\bfPhi_n(\bfeta) 
= 
\begin{pmatrix}
\frac{1}{N}\sum_{i=1}^{N}\Big[\frac{R_i(y_i-\mu)}{\pi_i^{\A}}+\Delta\frac{R_i-\pi_i^{\A}}{\pi_i^{\A}}\Big] \\
\frac{1}{N}\sum_{i=1}^{N}R_i\bfx_i-\frac{1}{N}\sum_{i \in \mS_{\B}}d_i^{\B}\pi_i^{\A}\bfx_i
\end{pmatrix}
={\bf 0}\,,
\end{equation}
where $\Delta = \mu$ if $\hat\mu = \hat\mu_{\IPWone}$ and $\Delta = 0$ if $\hat\mu = \hat\mu_{\IPWtwo}$. This formation is similar to the one used by Lunceford and Davidian (2004). Under the joint randomization of the propensity score model and the sampling design for $\mS_{\B}$, we have 
$E\{\bfPhi_n(\bfeta)\} = {\bf 0}$ when $\bfeta^{\intercal} =  \bfeta_0^{\intercal} = (\mu_y,\bftheta_0^{\intercal})$. Consistency of the estimator $\hat{\mathbf{\bfeta}}$ follows similar arguments in Section 3.2 of Tsiatis (2006). 

Under conditions {\bf C1}-{\bf C6}, we have $\bfPhi_n(\hat{\bfeta}) = {\bf 0}$ and $\bfPhi_n(\bfeta_0) = O_p(n_{\A}^{-1/2})$.  
By applying the first order Taylor expansion to $\bfPhi_n(\hat{\bfeta})$ around $\bfeta_0$, we further have 
\begin{equation}
\hat{\bfeta}-\bfeta_0 = \big[\bfphi_n(\bfeta_0)\big]^{-1}\bfPhi_n(\bfeta_0) + o_p(n_{\A}^{-1/2}) 
= \big[E\big\{\bfphi_n(\bfeta_0)\big\}\big]^{-1}\bfPhi_n(\bfeta_0) + o_p(n_{\A}^{-1/2}) \,,
\label{expansion}
\end{equation}
where $\bfphi_n(\bfeta) = \partial \bfPhi_n(\bfeta)/\partial \bfeta$ and is given by 
\[
\bfphi_n(\bfeta)  = \frac{1}{N}
\begin{pmatrix}
    -\sum_{i =1}^{N}\big\{(1-\frac{\Delta}{\mu})\frac{R_i}{\pi_{i}^{\A}}+\frac{\Delta}{\mu}\big\}    & -\sum_{i =1}^{N}\frac{1-\pi_{i}^{\A}}{{\pi_{i}^{\A}}}R_i(y_i-\mu+\Delta) \bfx_i^{\intercal}\\
    \bfzero &  -\sum_{i \in \mS_{\B}}d_i^{\B}\pi_{i}^{\A}(1-\pi_{i}^{\A})\bfx_i\bfx_i^{\intercal}
\end{pmatrix} \,.
\]
It follows that $\hat\mu = \mu_y +O_p(n_{\A}^{-1/2})$ and 
\[
Var(\hat\bfeta) = \big[E\big\{\bfphi_n(\bfeta_0)\big\}\big]^{-1}Var\big\{\bfPhi_n(\bfeta_0)\big\}\big[E\big\{\bfphi_n(\bfeta_0)\big\}^{\intercal}\big]^{-1} + o(n_{\A}^{-1}) \,.
\]
It can be shown that 
\[
 \big[E\big\{\bfphi_n(\bfeta)\big\}\big]^{-1} 
 =
 \begin{pmatrix}
    -1     & \frac{\Delta}{\mu}{\bf b}_1^{\intercal}+(1-\frac{\Delta}{\mu}){\bf b}_2^{\intercal}\\
   \mathbf{0} &  -[\frac{1}{N}\sum_{i =1}^{N}\pi_{i}^{\A}(1-\pi_{i}^{\A})\bfx_i\bfx_i^{\intercal}]^{-1}
   \end{pmatrix} \,,
\]
where the expressions for ${\bf b}_1$ and ${\bf b}_2$ are given in Theorem 1. 
The other major piece $Var\big\{\bfPhi_n(\bfeta_0)\big\}$ can be found by using the decomposition 
$\bfPhi_n(\bfeta)={\bf A}_1+ {\bf A}_2$ where 
\[
{\bf A}_1= \frac{1}{N}\sum_{i=1}^{N}
\begin{pmatrix}
\frac{R_i(y_i-\mu)}{\pi_{i}^{\A}}+\Delta \, \frac{R_i-{\pi_{i}^{\A}}}{\pi_{i}^{\A}}\\
R_i\bfx_i-{\pi_{i}^{\A}}\bfx_i
\end{pmatrix}\,, \;\;\;
{\bf A}_2=\frac{1}{N}
\begin{pmatrix} 0 \\ \sum_{i=1}^{N}\pi_{i}^{\A}\bfx_i-\sum_{i \in  S_B}d_i^{\B}\pi_{i}^{\A}\bfx_i
\end{pmatrix} \,.
\]
It follows that $Var\big\{\bfPhi_n(\bfeta_0)\big\} = {\bf V}_1 + {\bf V}_2$ where ${\bf V}_1 = Var({\bf A}_1)$ which only involves the propensity score model and 
${\bf V}_2 = Var({\bf A}_2)$ which only involves the sampling design for $\mS_{\B}$, both evaluated at $\bfeta = \bfeta_0$. We have 
\[
{\bf V}_1 = \frac{1}{N^2}\sum_{i =1}^{N}
\begin{pmatrix}
  \{(1-\pi_{i}^{\A})/\pi_{i}^{\A}\}(y_i-\mu+\Delta)^2   & {(1-\pi_{i}^{\A})(y_i-\mu+\Delta)\bfx_i}^{\intercal} \\
(1-\pi_{i}^{\A})(y_i-\mu+\Delta) \bfx_i & \pi_{i}^{\A}(1-\pi_{i}^{\A})\bfx_i\bfx_i^{\intercal}
\end{pmatrix}
\]
and 
\[
{\bf V}_2 = 
\begin{pmatrix} 0    &  \mathbf{0}^{\intercal} \\
\mathbf{0}  & {\bf D}
\end{pmatrix}\,,
\] 
where ${\bf D} = N^{-2}V_p\big(\sum_{i \in  \mS_{\B}}d_i^{\B}\pi_{i}^{\A}\bfx_i \big)$
is the design-based variance-covariance matrix under the probability sampling design for $\mS_{\B}$. The asymptotic variance for the IPW estimator 
$\hat\mu$ is obtained as the first diagonal element of the matrix 
$\big[E\big\{\bfphi_n(\bfeta_0)\big\}\big]^{-1}\big\{{\bf V}_1+{\bf V}_2\big\}\big[E\big\{\bfphi_n(\bfeta_0)\big\}^{\intercal}\big]^{-1}$ due to (\ref{expansion}).

\medskip

\no 
{\bf   A.3 \ Proof of Theorem 2}

\medskip

\no
The double robustness property is straightforward from the construction of the estimator. 
We first show that the estimation of the outcome regression model parameters $\bfbeta$ has no impact on the asymptotic variance of 
$\hat{\mu}_{\DRtwo}$. We assume that $\hat\bfbeta -\bfbeta^* = O_p(n_{\A}^{-1/2})$ for some fixed $\bfbeta^*$ regardless of the true regression model. 
Treating $\hat{\mu}_{\DRtwo}$ as a function of $\hat\bfbeta$ and making a Taylor expansion around $\bfbeta^*$ and under conditions {\bf C1}-{\bf C4}, we have 
\begin{eqnarray}
\hat{\mu}_{\DRtwo} &=& \frac{1}{\hat{N}^{\A}}\sum_{i =1}^{N}\frac{R_i\big\{y_i-m(\bfx_i,\bfbeta^{*})\big\}}{\hat{\pi}_i^{\A}}+\frac{1}{\hat{N}^{\B} }\sum_{i \in \mS_{\B}}d_i^{\B}m(\bfx_i,\bfbeta^{*}) \label{ae1} \\
&+&\bigg\{\frac{1}{\hat{N}^{\B} }\sum_{i \in \mS_{\B}}d_i^{\B}\dot{m}(\bfx_i,\bfbeta^{*})-\frac{1}{\hat{N}^{\A}}\sum_{i =1}^{N}\frac{R_i\dot{m}(\bfx_i,\bfbeta^{*})}{\hat{\pi}_i^{\A}}\bigg\}(\hat{\bfbeta}-\bfbeta^{*}) \nonumber\\
&+& O_p(n_{\A}^{-1}) \,,
\nonumber
\end{eqnarray}
where   $\dot{m}(\bfx,\bfbeta)={\partial}m(\bfx,\bfbeta)/{\partial \bfbeta}$. Under conditions {\bf C2} and {\bf C4}, we have 
\begin{eqnarray*}
\frac{1}{\hat{N}^{\B} }\sum_{i \in \mS_{\B}}d_i^{\B}\dot{m}(\bfx_i,\bfbeta^{*})-\frac{1}{N}\sum_{i =1}^{N}\dot{m}(\bfx_i,\bfbeta^{*}) &=& o_p(1) \,, \\
\frac{1}{\hat{N}^{\A}}\sum_{i =1}^{N}\frac{R_i\dot{m}(\bfx_i,\bfbeta^{*})}{\hat{\pi}_i^{\A}}-\frac{1}{N}\sum_{i =1}^{N}\dot{m}(\bfx_i,\bfbeta^{*}) &=& o_p(1) \,,
\end{eqnarray*}
which implies that $(\hat{N}^{\B})^{-1}\sum_{i \in \mS_{\B}}d_i^{\B}\dot{m}(\bfx_i,\bfbeta^{*})-(\hat{N}^{\A})^{-1}\sum_{i =1}^{N}R_i\dot{m}(\bfx_i,\bfbeta^{*})/\hat{\pi}_i^{\A}=o_p(1)$ and 
\begin{equation}
\hat{\mu}_{\DRtwo} =\frac{1}{\hat{N}^{\A}}\sum_{i =1}^{N}\frac{R_i\big\{y_i-m(\bfx_i,\bfbeta^{*})\big\}}{\hat{\pi}_i^{\A}}+\frac{1}{\hat{N}^{\B} }\sum_{i \in \mS_{\B}}d_i^{\B}m(\bfx_i,\bfbeta^{*})+o_p\big(n_{\A}^{-1/2}\big)\,. 
\label{dr3}
\end{equation}

We now derive the asymptotic variance of $\hat{\mu}_{\DRtwo}$ under the propensity score model and the sampling design for $\mS_{\B}$. The first part of 
$\hat{\mu}_{\DRtwo}$ given in (\ref{dr3}) is the IPW estimator $\hat{\mu}_{\IPWtwo}$ given in (\ref{dr1}) with $y_i$ replaced by $y_i - m(\bfx_i,\bfbeta^*)$. Using the asymptotic expansion developed in (\ref{expansion}) on $\hat{\mu}_{\IPWtwo}$, we have 
\begin{eqnarray*}
\frac{1}{\hat{N}^{\A}}\sum_{i =1}^{N}\frac{R_i\big\{y_i-m(\bfx_i,\bfbeta^{*})\big\}}{\hat{\pi}_i^{\A}}
&=&h_{\N}+\frac{1}{N}\sum_{i =1}^{N} R_i\bigg\{\frac{y_i-m(\bfx_i,\bfbeta^{*})-h_{\N}}{\pi_{i}^{\A}}-{\bf b}_3^{\intercal}\bfx_i\bigg\}\\
& & + \; {\bf b}_3^{\intercal}\frac{1}{N}\sum_{i \in \mS_{\B}}d_i^{\B}\pi_{i}^{\A}\bfx_i+o_p(n_{\A}^{-1/2})\,,
\end{eqnarray*}
where $h_{\N}=N^{-1}\sum_{i =1}^{N}\big\{y_i - m(\bfx_i,\bfbeta^{*})\big\}$ and 
\[
{\bf b}_3^{\intercal} = {\bigg[\frac{1}{N}\sum_{i =1}^{N}(1-\pi_{i}^{\A})\big\{y_i - m(\bfx_i,\bfbeta^{*})-h_{\N}\big\}\bfx_i^{\intercal}\bigg]}\bigg\{\frac{1}{N}\sum_{i =1}^{N}\pi_{i}^{\A}(1-\pi_{i}^{\A})\bfx_i\bfx_i^{\intercal}\bigg\}^{-1}\,.
\]
The second part of $\hat{\mu}_{\DRtwo}$ given in (\ref{dr3}) is the H\'ajek estimator under the probability sampling design for $\mS_{\B}$, which has the following expansion 
\[
\frac{1}{\hat{N}^{\B} }\sum_{i \in \mS_{\B}}d_i^{\B}m_i=\frac{1}{N}\sum_{i =1}^{N}m_i+\frac{1}{N}\sum_{i \in \mS_{\B}}d_i^{\B}\Big\{m_i-\frac{1}{N}\sum_{i =1}^{N}m_i\Big\}+O_p(n_{\B}^{-1})\,,
\]
where $m_i = m(\bfx_i,\bfbeta^{*})$. Putting the two parts together leads to 
\[
\hat{\mu}_{\DRtwo} - \mu_y = \frac{1}{N}\sum_{i =1}^{N} R_i\bigg\{\frac{y_i-m(\bfx_i,\bfbeta^{*})-h_{\N}}{\pi_{i}^{\A}}-{\bf b}_3^{\intercal}\bfx_i\bigg\} 
+\frac{1}{N}\sum_{i \in \mS_{\B}}d_i^{\B} t_i + o_p(n_{\A}^{-1/2})\,,
\]
where $t_i=\pi_{i}^{\A}\bfx_i^{\intercal} {\bf b}_3 + m(\bfx_i,\bfbeta^{*}) -N^{-1}\sum_{i =1}^{N}m(\bfx_i,\bfbeta^{*})$. It follows that the asymptotic variance of $\hat{\mu}_{\DRtwo}$ is given by $V_{\DRtwo}$ as specified in Theorem 2.

\bigskip

\bigskip

\no {\bf REFERENCES}

\bigskip

\def\beginref{\begingroup
                \clubpenalty=10000
                \widowpenalty=10000
                \normalbaselines\parindent 0pt
                \parskip.0\baselineskip
                \everypar{\hangindent1em}}
\def\endref{\par\endgroup}
\renewcommand{\baselinestretch}{1}

\beginref


Baker, R., Brick, J. M., Bates, N. A., Battaglia, M., Couper, M. P., Dever, J. A., Gile, K. J. and Tourangeau, R. (2013), `` Report of the AAPOR Task Force on Non-probability Sampling,''  {\em Journal of Survey Statistics and Methodology}, 1, 90-143.




Brick, J.M. (2015), ``Compositional Model Inference'',  {\em Proceedings of the Survey Research Methods Section}, Joint Statistical Meetings, American Statistical Association, Alexandria, VA, 299-307.

Cao, W., Tsiatis, A.A., and Davidian, M. (2009), ``Improving Efficiency and Robustness of the Doubly Robust Estimator for a Population Mean with Incomplete Data,'' {\em Biometrika}, 96,  723-734.


Citro, C.F. (2014), ``From Multiple Modes for Surveys to Multiple Data Sources for Estimates,'' {\em Survey Methodology}, 40, 137-161.







Hansen, M.H. (1987), ``Some History and Reminiscences on Survey Sampling,'' {\em Statistical Science}, 2, 180-190.

Haziza, D., and Rao, J.N.K. (2006), ``A Nonresponse Model Approach to Inference under Imputation for Missing Survey Data,'' {\em Survey Methodology}, 32, 53-64.

Horvitz, D.G., and Thompson, D.J. (1952), ``A Generalization of Sampling Without Replacement From a Finite Universe,'' {\em Journal of the American Statistical Association}, 47, 663- 685.


Kang, J.D.Y., and Schafer, J.L. (2007), ``Demystifying Double Robustness: A Comparison of Alternative Strategies for Estimating a Population Mean from Incomplete Data, {\em Statistical Science}, 22, 523-539.

Kim, J.K., and Haziza, D. (2014), ``Doubly Robust Inference with Missing Data in Survey Sampling,'' {\em Statistica Sinica}, 24, 375-394.

Kim, J.K., and Park, H.A. (2006), ``Imputation Using Response Probability,'' {\em The Canadian Journal of Statistics}, 34, 171-182.


Kott, P.S. (1994), ``A Note on Handling Nonresponse in Sample Surveys,'' {\em Journal of the American Statistical Association}, 89, 693-696.

Lee, S. (2006). ``Propensity Score Adjustment as a Weighting Scheme for Volunteer Panel Web Surveys,'' {\em Journal of Official Statistics}, 22, 329-349.

Lee, S., and Valliant, R. (2009), ``Estimation for Volunteer Panel Web Surveys Using Propensity Score Adjustment and Calibration Adjustment,'' {\em Sociological Methods $\&$ Research}, 37, 319-343.



Little, R. J., and Rubin, D.B. (2002), ``{\em Statistical Analysis with Missing Data},'' Wiley, New York, Second Edition.

Lunceford, J. K., and Davidian, M. (2004), ``Stratification and Weighting Via the Propensity Score in Estimation of Causal Treatment Effects: A Comparative Study,'' {\em Statistics in Medicine}, 23, 2937-2960.

Neyman, J. (1938). ``Contribution to the Theory of Sampling Human Populations'', {\em Journal of the American Statistical Association}, 33, 101-116.

Rao, J.N.K.(2005), ``Interplay Between Sample Survey Theory and Practice: An Appraisal'', {\em Survey Methodology}, 31, 117-138.

Rao, J.N.K. and Fuller, W.A. (2017), ``Sample Survey Theory and Methods: Past, Present, and Future Directions'', {\em Survey Methodology}, 43, 145-160.  

Rivers, D. (2007), ``Sampling for Web Surveys'', {\em Proceedings of the Survey Research Methods Section}, Joint Statistical Meetings, American Statistical Association, Alexandria, VA, 1-26.

Robins, J.M., Rotnitzky, A., and Zhao. L.P. (1994), ``Estimation of Regression Coefficients When Some Regressors Are Not Always Observed'', {\em Journal of the American Statistical Association}, 89, 846-866.

Rosenbaum, P.R., and Rubin, D.B. (1983), ``The Central Role of the Propensity Score in Observational Studies for Causal Effects,'' {\em Biometrika}, 70, 41-55.


 
Rubin, D.B. (1976), ``Inference and Missing Data,'' {\em Biometrika},  63,  581-592.




Scharfstein, D.O., Rotnitzky, A., and Robins, J.M. (1999), ``Adjusting for Nonignorable Drop-Out Using Semiparametric Nonresponse Models,'' {\em Journal of the American Statistical Association}, 94, 1096-1120.


Tan, Z. (2007), ``Comment: Understanding OR, PS and DR,'' {\em Statistical Science}, 22, 560-568.

 
 
Tourangeau, R.,  Conrad, F.G., and Couper, M.P. (2013), ``{\em The Science of Web Surveys},'' Oxford University Press, First Edition.
 
 Tsiatis, A. A. (2006), ``{\em Semiparametric Theory and Missing Data}'', Springer, New York. 
 
 Valliant, R., and Dever J. A. (2011), ``Estimating Propensity Adjustments for Volunteer Web Surveys,''
{\em Sociological Methods $\&$ Research}, 40, 105-137.

Vavreck, L., and Rivers, D. (2008), ``The 2006 Cooperative Congressional Election Study,'' {\em Journal of Elections, Public Opinion and Parties}, 18, 355-366.

Wu, C., and Sitter, R. R. (2001), ``A Model-Calibration Approach to Using Complete Auxiliary Information from Survey Data'', {\em Journal of the American Statistical Association}, 96, 185-193.

\newpage

\begin{table}[!htbp]
\centering
\caption {Simulated RB$\%$ and MSE of Estimators of $\mu$ ($n_{\A}=500$, $n_{\B}=1000$)}  
\label{tab1} 
\begin{tabular}{llrrrrrrrr}
   \hline
& &\multicolumn{2}{c}{$\rho=0.30$} & \multicolumn{2}{c}{$\rho=0.50$} & \multicolumn{2}{c}{$\rho=0.80$}\\
Models  &Estimator & $\%$RB  & MSE &  $\%$RB  & MSE &  $\%$RB  & MSE\\ 
  \hline
TT & $\hat{\mu}_{\A}$ & 28.27 & 7.01 & 28.16 & 6.84 & 28.08 & 6.78 \\
&$\hat{\mu}_{\Cone}$ & -91.21 & 70.49 & -91.06 & 70.56 & -90.88 & 70.49\\
&$\hat{\mu}_{\Ctwo}$ & 14.23 & 1.99 & 15.49 & 2.14 & 16.84 & 2.46 \\
&$\hat{\mu}_{\IPWone}$ & 0.49 & 0.41 & 0.49 & 0.20 & 0.48 & 0.13 \\
&$\hat{\mu}_{\IPWtwo}$ & -0.17 & 0.33 & -0.16 & 0.12 & -0.16 & 0.05 \\
&$\hat{\mu}_{\REG}$ & 0.13 & 0.29 & 0.09 & 0.10 & 0.06 & 0.03  \\
&$\hat{\mu}_{\DRtwo}$ & 0.05 & 0.31 &0.04 & 0.10 & 0.04 & 0.03 \\
 \hline
  FT & $\hat{\mu}_{\A}$ & 28.27 & 7.01 &28.16 & 6.84 & 28.08 & 6.78 \\
  &$\hat{\mu}_{\Cone}$ & -91.21 & 70.49 & -91.06 & 70.56 & -90.88 & 70.49\\
&$\hat{\mu}_{\Ctwo}$ & 14.23 & 1.99 & 15.49 & 2.14 & 16.84 & 2.46 \\
&$\hat{\mu}_{\IPWone}$ & 0.49 & 0.41 & 0.49 & 0.20 & 0.48 & 0.13 \\
&$\hat{\mu}_{\IPWtwo}$ & -0.17 & 0.33 & -0.16 & 0.12 & -0.16 & 0.05 \\
&$\hat{\mu}_{\REG}$ & 24.53 & 5.35 & 24.50 & 5.21 & 24.48 & 5.16  \\
&$\hat{\mu}_{\DRtwo}$ & -0.05 & 0.33 & -0.05 & 0.12& -0.05 & 0.05 \\
    \hline
   TF &$\hat{\mu}_{\A}$ & 28.27 & 7.01 & 28.16 & 6.84 & 28.08 & 6.78 \\
   &$\hat{\mu}_{\Cone}$ & -89.91 & 68.49 & -89.75 & 68.55 & -89.55 & 68.44 \\
&$\hat{\mu}_{\Ctwo}$ &  30.80 & 8.29 & 31.83 & 8.73 & 32.98 & 9.35 \\
&$\hat{\mu}_{\IPWone}$ & 24.61 & 5.52 & 24.57 & 5.38  & 24.53 & 5.33 \\
&$\hat{\mu}_{\IPWtwo}$ & 24.56 & 5.36 & 24.51 & 5.21&24.48 & 5.16 \\
&$\hat{\mu}_{\REG}$ & 0.13 & 0.29  & 0.09 & 0.10 & 0.06 & 0.03  \\
&$\hat{\mu}_{\DRtwo}$ & 0.18 & 0.29 & 0.12 & 0.09 & 0.07 & 0.03 \\
\hline
\end{tabular}
\end{table}

\newpage

\begin{table}[!htbp]
\centering
\caption {Simulated $\%$ RB and $\%$CP of Variance Estimators ($n_{\A}=500$, $n_{\B}=1000$)} 
\label{tab2} 
\begin{tabular}{llrrrrrrrr}
   \hline
& &\multicolumn{2}{c}{$\rho=0.30$} & \multicolumn{2}{c}{$\rho=0.50$} & \multicolumn{2}{c}{$\rho=0.80$}\\
Models  &Estimator & $\%$RB  & \%CP &  $\%$RB  & \%CP &  $\%$RB  & \%CP\\ 
  \hline
TT & $v_{\IPWone}$ & 1.03 & 95.18 & 0.84 & 95.06  & 0.09 & 94.64\\
& $v_{\IPWtwo}$ & 2.82 & 94.97 & 2.21 & 94.96 & 0.58 & 94.73\\
& $v_{\PLUG}$ & 2.81 & 94.95 & 2.45 & 95.08 & 1.50 & 94.87 \\
&$v_{\KH}$ & -0.18 & 94.97 &  -0.41 & 94.65 & -1.37 & 94.73 \\
 \hline
 FT & $v_{\IPWone}$ & 1.03 & 95.18 & 0.84 & 95.06 & 0.09 & 94.64\\
& $v_{\IPWtwo}$ & 2.82 & 94.97 & 2.21 & 94.96  & 0.58 & 94.73\\
 & $v_{\PLUG}$ & 2.64 & 94.87 & 1.60 & 95.06  & -0.91 & 94.53 \\
&$v_{\KH}$ & 1.93 & 95.21&4.01 & 95.21 & 5.80 & 95.47\\
    \hline
  TF & $v_{\IPWone}$ & 1.47 & 3.35 & 1.51 & 0.14 & 0.84 & 0.02\\
& $v_{\IPWtwo}$ & 2.97 & 0.62  & 5.03 & 0.00 & 8.75 & 0.00\\
   & $v_{\PLUG}$ &-20.62 & 91.74 & -18.81 & 91.92 & -11.93 & 93.53\\
  &$v_{\KH}$ & -17.12 & 92.42 & -10.94 & 93.21 &-4.47 &94.34 \\
\hline
\end{tabular}
\end{table}

\newpage

\begin{table}[!htbp]
\singlespacing
\centering
\caption {Marginal Distributions of Common Covariates from the Three Samples} 
\label{tab3} 
\begin{tabular}{lllll}
\hline
& & $\hat{X}_{\PRC}$ &$\hat{X}_{\BRFSS}$  & $\hat{X}_{\CPS}$ \\
  \hline
   Age category & $<$30 & 18.29 & 20.91  & 21.24\\ 
  & $>$=30,$<$50 & 32.60 & 33.28 & 33.55\\ 
  & $>$=50,$<$70 & 38.68 & 32.69 & 32.60 \\ 
  & $>$=70 & 10.43 & 13.12 &12.62 \\ 
   \hline
 Gender & Female & 54.36 & 51.32 &51.83 \\ 
   \hline
    Race&  White only & 82.28 & 75.05 &78.61\\ 
   \hline
   Race& Black only&  8.83 & 12.59 & 12.46\\ 
   \hline
   Origin& Hispanic/Latino &  9.27 & 16.52  &15.60\\ 
   \hline
      Region & Northeast & 19.96 & 17.70&18.00\\ 
   \hline
   Region & South & 27.53 & 38.27& 37.34\\ 
   \hline
   Region & West & 29.87 & 23.18&23.47 \\ 
   \hline
   Marital status& Married &   50.35 & 50.82& 52.82\\ 
         \hline
  Employment & Working &  52.13 & 56.63  & 58.90\\ 
   \hline
   Employment& Retired & 24.34 & 17.93 & 14.34\\ 
     \hline
      Education& High school or less & 21.63 & 42.66& 40.65 \\ 
       \hline
    Education   & Bachelor's degree and above & 41.57 & 26.32 &30.90 \\  
   \hline
     Education  & Bachelor's degree &  22.07 & NA &19.83 \\ 
   \hline
        Education & Postgraduate &  19.50 &  NA& 11.07\\ 
   \hline
  Household & Presence of child in household&  28.93 & 36.78 &NA\\ 
   \hline
   Household & Home ownership & 65.42 & 67.19 &NA\\ 
   \hline
       Health& Smoke everyday &15.74 & 11.49&NA\\ 
   \hline
   Health& Smoke never & 79.80 & 83.28  &NA\\ 
   \hline
    Financial status& No money to see doctors & 20.68 & 13.27 &NA\\ 
   \hline
   Financial status& Having medical insurance & 89.15 & 87.83 &NA\\ 
   \hline
 Financial status & Household income $<$ 20K & 16.14 &NA&15.32\\ 
   \hline
  Financial status & Household income $>$100K & 19.89 & NA& 23.32 \\ 
   \hline
    Volunteer works & Volunteered & 50.98 & NA&24.83\\ 
   \hline
    \end{tabular}
\end{table}

\begin{table}[!htbp]
\centering
\caption {Estimated Population Mean of  $y$ Using A Single Set of Common Covariates} 
\label{tab4} 
\begin{tabular}{llllllllll}
\hline
 Response Variable $y$ && &$\hat{\mu}_{\A}$&$\hat{\mu}_{\Cone}$&$\hat{\mu}_{\Ctwo}$&$\hat{\mu}_{\IPWone}$&$\hat{\mu}_{\IPWtwo}$&$\hat{\mu}_{\REG}$&$\hat{\mu}_{\DRtwo}$\\
  \hline
 \%Talked with & BRFSS &  & \multicolumn{1}{c}{\multirow{2}{*}{46.13}} & 0.08 & 46.87 & 44.68 & 45.53 & 45.72 & 45.72 \\ 
  neighbours frequently &  CPS & \multicolumn{1}{c}{} & \multicolumn{1}{c}{} &0.02 & 45.60 & 45.11 & 45.41 & 45.76 & 45.79 \\
 \hline
    \%Tended to trust & BRFSS &  &  \multicolumn{1}{c}{\multirow{2}{*}{58.97}} & 0.10 & 60.74 & 54.62 & 55.69 & 55.30 & 55.37 \\ 
   neighbours& CPS&\multicolumn{1}{c}{}  &\multicolumn{1}{c}{}  & 0.02 & 56.89 & 55.14 & 55.53 & 55.66 & 55.71\\
   \hline
    \%Expressed opinions&BRFSS &  & \multicolumn{1}{c}{\multirow{2}{*}{26.54}}  &0.04 & 25.90 & 23.78 & 24.23 & 23.98 & 24.17  \\ 
   at a government level& CPS&\multicolumn{1}{c}{} &\multicolumn{1}{c}{} & 0.01 & 24.62 & 24.25 & 24.40 & 24.31 & 24.52\\
    \hline
      \%Voted local &BRFSS &   & \multicolumn{1}{c}{\multirow{2}{*}{74.97 }}  & 0.13 & 77.00 & 69.94 & 71.28 & 70.70 & 70.87\\ 
     elections& CPS&\multicolumn{1}{c}{} &  \multicolumn{1}{c}{} &0.03 & 72.57 & 70.92 & 71.38 & 71.59 & 71.69 \\
     \hline 
\%Participated in&BRFSS  &  & \multicolumn{1}{c}{\multirow{2}{*}{20.97}}   & 0.03 & 18.48 & 19.80 & 20.18 & 20.02 & 20.16  \\ 
   school groups & CPS&\multicolumn{1}{c}{} & \multicolumn{1}{c}{} &0.01 & 19.97 & 20.59 & 20.73 & 20.55 & 20.71\\
 \hline
    \%Participated in&BRFSS &    & \multicolumn{1}{c}{\multirow{2}{*}{14.11}}  &  0.02 & 13.34 & 13.01 & 13.26 & 13.28 & 13.29 \\ 
   service organizations  &CPS& \multicolumn{1}{c}{} & \multicolumn{1}{c}{} & 0.00 & 13.10 & 13.42 & 13.51 & 13.50 & 13.56 \\
   \hline
   Days had at  least & BRFSS &   & \multicolumn{1}{c}{\multirow{2}{*}{ 5.30 }} &0.01 & 5.26 & 4.86 & 4.95 & 4.93 & 4.97 \\ 
   one drink last month& CPS&\multicolumn{1}{c}{} &\multicolumn{1}{c}{} & 0.00 & 5.04 & 4.92 & 4.95 & 4.99 & 5.01 \\
   \hline
 \end{tabular}
\end{table}

\begin{table}[!htbp]
\centering
\caption {Estimated Population Mean of  $y$ Using Separate Sets of Common Covariates} 
\label{tab5} 
\begin{tabular}{llllllllll}
\hline
 Response Variable $y$ && &$\hat{\mu}_{\A}$&$\hat{\mu}_{\Cone}$&$\hat{\mu}_{\Ctwo}$&$\hat{\mu}_{\IPWone}$&$\hat{\mu}_{\IPWtwo}$&$\hat{\mu}_{\REG}$&$\hat{\mu}_{\DRtwo}$\\
  \hline
  \%Talked with & BRFSS &  & \multicolumn{1}{c}{\multirow{2}{*}{46.13}} & 0.08 & 46.23 & 43.54 & 44.23 & 44.64 & 44.63 \\ 
 neighbours frequently &  CPS & \multicolumn{1}{c}{} & \multicolumn{1}{c}{} & 0.01 & 41.34 & 39.49 & 40.32 & 40.43 & 40.14\\
 \hline
    \%Tended to trust & BRFSS &  &  \multicolumn{1}{c}{\multirow{2}{*}{58.97}} & 0.10 & 60.86 & 55.17 & 56.07 & 56.06 & 56.04 \\ 
   neighbours& CPS&\multicolumn{1}{c}{}  &\multicolumn{1}{c}{}  & 0.02 & 55.40 & 52.75 & 53.92 & 52.97 & 53.12\\
   \hline
    \%Expressed opinions&BRFSS &   & \multicolumn{1}{c}{\multirow{2}{*}{26.54}}  & 0.04 & 24.56 & 22.03 & 22.38 & 22.33 & 22.37 \\ 
   at a government level& CPS&\multicolumn{1}{c}{} &\multicolumn{1}{c}{} & 0.01 & 21.02 & 19.76 & 20.17 & 19.91 & 19.91\\
    \hline
      \%Voted local &BRFSS &  & \multicolumn{1}{c}{\multirow{2}{*}{74.97 }}  &0.13 & 77.34 & 70.68 & 71.84 & 71.51 & 71.77  \\ 
     elections& CPS&\multicolumn{1}{c}{} &  \multicolumn{1}{c}{} & 0.03 & 70.64 & 67.99 & 69.43 & 68.10 & 68.26\\
     \hline 
\%Participated in&BRFSS  &  & \multicolumn{1}{c}{\multirow{2}{*}{20.97}}   & 0.03 & 18.14 & 19.37 & 19.68 & 19.79 & 19.75  \\ 
   school groups & CPS&\multicolumn{1}{c}{} & \multicolumn{1}{c}{} &0.01 & 14.16 & 13.46 & 13.74 & 13.32 & 13.12\\
 \hline
    \%Participated in&BRFSS &  & \multicolumn{1}{c}{\multirow{2}{*}{14.11}}  & 0.02 & 12.54 & 11.85 & 12.04 & 12.09 & 11.98  \\ 
   service organizations  &CPS& \multicolumn{1}{c}{} & \multicolumn{1}{c}{} &  0.00 & 9.30 & 8.79 & 8.98 & 8.68 & 8.54\\
   \hline
   Days had at  least& BRFSS &   & \multicolumn{1}{c}{\multirow{2}{*}{ 5.30 }}& 0.01 & 5.11 & 4.71 & 4.78 & 4.81 & 4.82 \\ 
   one drink last month & CPS&\multicolumn{1}{c}{} &\multicolumn{1}{c}{} & 0.00 & 5.15 & 4.96 & 5.07 & 5.06 & 5.09\\
   \hline
 \end{tabular}
\end{table}

\end{document}